\documentclass[aps,twocolumn,showpacs]{revtex4}
\pdfoutput=1
\usepackage{amsmath}
\usepackage{graphicx}
\usepackage{bm}
\usepackage{color}

\newcommand{\Cca}{\mathcal{C}}

\begin{document}

\title{Electrically detected interferometry of Majorana fermions in a topological insulator}

\author{A. R. Akhmerov}
\affiliation{Instituut-Lorentz, Universiteit Leiden,
P.O. Box 9506, 2300 RA Leiden, The Netherlands}

\author{Johan Nilsson}
\affiliation{Instituut-Lorentz, Universiteit Leiden,
P.O. Box 9506, 2300 RA Leiden, The Netherlands}

\author{C. W. J. Beenakker}
\affiliation{Instituut-Lorentz, Universiteit Leiden,
P.O. Box 9506, 2300 RA Leiden, The Netherlands}

\date{March 2009}
\begin{abstract}
We show how a chiral Dirac fermion (a massless electron or hole) can be converted into a pair of neutral chiral Majorana fermions (a particle equal to its own antiparticle). These two types of fermions exist on the metallic surface of a topological insulator, respectively, at a magnetic domain wall and at a magnet-superconductor interface. Interferometry of Majorana fermions is a key operation in topological quantum computation, but the detection is problematic since these particles have no charge. The Dirac-Majorana converter enables electrical detection of the interferometric signal.
\end{abstract}
\pacs{74.45.+c, 03.67.Lx, 71.10.Pm, 73.23.-b}

\maketitle
There is growing experimental evidence \cite{Rad08,Dol08,Wil09} that the $5/2$ fractional quantum Hall effect (FQHE) is described by the Moore-Read state \cite{Moo91}. This state has received much interest in the context of quantum computation \cite{Nay08}, because its quasiparticle excitations are Majorana bound states. A qubit can be stored nonlocally in a pair of widely separated Majorana bound states, so that no local source of decoherence can affect it \cite{Kit01}. The state of the qubit can be read out and changed in a fault-tolerant way by edge state interferometry \cite{Das05,Ste06,Bon06}. This ``measurement based topological quantum computation'' \cite{Bon08} combines static quasiparticles within the Hall bar to store the qubits, with mobile quasiparticles at the edge of the Hall bar to perform logical operations by means of interferometric measurements.

The electronic correlations in the Moore-Read state involve a pairing of spin-polarized fermions, equivalent to a superconducting pairing with  $p_{x}+ ip_{y}$ orbital symmetry \cite{Gre92,Rea00,Iva01}. Such an exotic pairing might occur naturally in the ${\rm Sr}_{2}{\rm RuO}_{4}$ superconductor \cite{Sar06}, or it might be produced artificially in $p$-wave superfluids formed by fermionic cold atoms \cite{Tew07}. Recently, Fu and Kane \cite{Fu08} showed how a conventional $s$-wave superconductor might produce Majorana bound states, if brought in proximity to a topological insulator. This class of insulators has metallic surface states with massless quasiparticles, as has been demonstrated in ${\rm Bi}_{x}{\rm Sb}_{1-x}$ alloys \cite{Hsi08} and ${\rm Bi}_{2}{\rm Se}_{3}$ single crystals \cite{Zha08,Xia08}. The latter material is particularly promising for applications because it remains a topological insulator at room temperature. The $5/2$ FQHE, in contrast, persists only at temperatures well below $1\,{\rm K}$ \cite{Rad08,Dol08,Wil09}.

While induced superconductivity in a topological insulator seems an attractive alternative to the FQHE for the purpose of quantum computation, one crucial difference creates a major obstacle: Quasiparticle excitations carry a charge in the FQHE, but they are charge-neutral in a superconductor. All known schemes \cite{Das05,Ste06,Bon06} for edge state interferometry rely on electrical detection, and this seems impossible if the edge states carry no electrical current. It is the purpose of this work to propose a way around this obstacle, by showing how a pair of neutral Majorana fermions can be converted phase coherently  and with unit probability into a charged Dirac fermion. 

 \begin{figure}
 \includegraphics[width=0.9\linewidth]{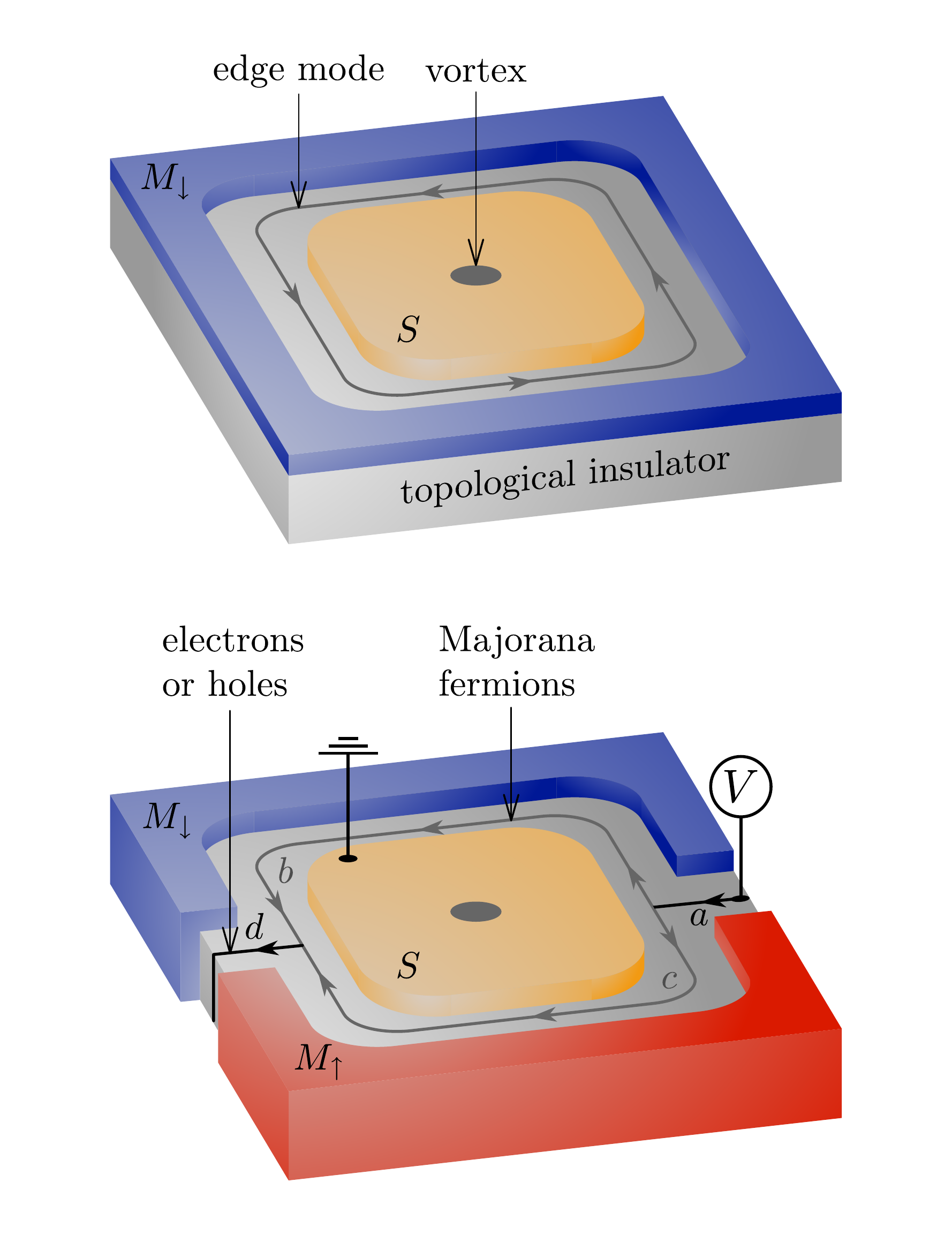}
 \caption{\label{fig:setup_3d} Three-dimensional topological insulator in proximity to ferromagnets with opposite polarization ($M_\uparrow$ and $M_\downarrow$) and to a superconductor ($S$). The top panel shows a single chiral Majorana mode along the edge between superconductor and ferromagnet. This mode is charge neutral, so it cannot be detected electrically. The Mach-Zehnder interferometer in the bottom panel converts a charged current along the domain wall into a neutral current along the superconductor (and vice versa). This allows for electrical detection of the parity of the number of enclosed vortices, as explained in the text.}
\end{figure}

We first give a qualitative description of the mechanism of electrically detected Majorana interferometry, and then present a quantitative theory. Our key idea is to combine edge channels of opposite chiralities in a single interferometer, by means of a magnetic domain wall.  The appearance of counterpropagating edge channels in a single superconducting domain is a special feature of a topological insulator in proximity to a ferromagnet, where the  propagation direction is determined by the way  time reversal symmetry is broken outside of the condensate (hence by the polarization of the ferromagnets) --- rather than being  determined by the order parameter  of the condensate  (as in  a $p_x\pm i p_y$ superconductor or FQHE droplet). 

Refering to the lower panel of Fig. \ref{fig:setup_3d}, we see that electrons or holes (with Dirac fermion operators $c^{\dagger}_{a}$ and $c_{a}$) propagate along the domain wall $a$ until they reach the superconductor, where they are split into a pair of Majorana fermions  $\gamma_{b}$ and $\gamma_{c}$ of opposite chirality:
\begin{equation}
 c^\dagger_a\rightarrow \gamma_b + i\gamma_c,\;\; c_a\rightarrow \gamma_b - i\gamma_c. \label{DtoM}
\end{equation}
(Here we have used that  $\gamma=\gamma^{\dagger}$, which is the defining property of a Majorana fermion.) 

The Dirac-to-Majorana fermion conversion expressed by Eq.\ \eqref{DtoM} relies on the fact that the electron or hole mode at the domain wall couples to \textit{a pair} of Majorana modes, so that the full information encoded by the complex fermion $c_{a}$ is encoded by two real fermions $\gamma_{b}$ and $\gamma_{c}$. This is the essential distinction from the process of electron tunneling into a Majorana bound state \cite{Sem07,Bol07,Tew08,Nil08}, which couples to \textit{a single} Majorana fermion and can therefore not transfer the full information.

Upon leaving the superconductor the Majorana fermions recombine into an electron $c_{d}^{\dagger}$ or hole $c_{d}$ depending on the number $n_{v}$ of superconducting vortices enclosed by the two arms of the interferometer,
\begin{equation}
\gamma_b + (-1)^{n_{v}}i\gamma_c \rightarrow c^\dagger_d,\;\;
\gamma_b - (-1)^{n_{v}}i\gamma_c \rightarrow c_d.
\end{equation}
For $n_{v}$ an even  integer, no charge is transfered to the superconductor, while for $n_{v}$ odd a charge $\pm 2e$ is absorbed by the superconducting condensate. The conductance $G$, measured by application of a voltage between a point on the domain wall and the superconductor, becomes equal (in the zero-temperature, zero-voltage limit) to $G=0$ for $n_{v}={\rm even}$ and $G=2e^{2}/h$ for $n_{v}={\rm odd}$. 

Proceeding now to a theoretical description, we recall that the surface of a three-dimensional topological insulator, in the presence of a  magnetization $\bm{M}(\bm{r})$ and superconducting order parameter $\Delta(\bm{r})$, is described  by the following Hamiltonian \cite{Fu08}:
\begin{equation}
H=\begin{pmatrix}
\bm{M}\cdot\bm{\sigma}+v_{F} \bm{p} \cdot \bm{\sigma}-E_{F}&\Delta \\
\Delta^{\ast}&\bm{M}\cdot\bm{\sigma}-v_{F} \bm{p} \cdot \bm{\sigma}+E_{F}
\end{pmatrix}.
\label{eq:HDBdG3d}
\end{equation}
Here $\bm{p}= (p_x,p_y,0)$ is the momentum on the surface, $\bm{\sigma}=(\sigma_{x},\sigma_{y},\sigma_z)$ is the vector of Pauli matrices, $v_{F}$ is the Fermi velocity, and $E_F$ the Fermi energy.  The two magnetizations $M_{\uparrow}$ and $M_{\downarrow}$ in Fig.~\ref{fig:setup_3d} correspond to ${\bm M}=(0,0,M_{0})$ and ${\bm M}=(0,0,-M_{0})$, respectively. Particle-hole symmetry is  expressed by the anticommutation $ H\Xi=-\Xi H$ of  the Hamiltonian with the operator 
\begin{equation}
\Xi=\begin{pmatrix}
0&i\sigma_{y}\Cca\\
-i\sigma_{y}\Cca&0
\end{pmatrix},
\end{equation}
with $\Cca$ the operator of complex conjuation.

There is a single chiral Majorana mode  with amplitude $\psi$  (group velocity $v_{m}$) at a boundary between a region with a superconducting gap and a region with a magnetic gap \cite{Fu08}. At a domain wall between two regions with opposite signs of  $M_z$ there are two chiral Dirac fermion modes, an electron mode  with amplitude $\phi^{e}$ and a hole mode  with ampitude $\phi^{h}$. The scattering matrix $S_{\textrm{in}}(\varepsilon)$ describes scattering  at excitation energy $\varepsilon$ from electron and hole modes (along edge $a$) to two Majorana modes (along edges $b$ and $c$  in Fig.~\ref{fig:setup_3d}), according to 
\begin{equation}
\begin{pmatrix}
 \psi_b\\
\psi_c
\end{pmatrix}=S_{\textrm{in}}
\begin{pmatrix}
 \phi_a^e\\
\phi_a^h
\end{pmatrix}.
\end{equation}

 Particle-hole symmetry for the scattering matrix is expressed by 
\begin{equation}
 S_{\textrm{in}}(\varepsilon)=S_{\textrm{in}}^*(-\varepsilon)
\begin{pmatrix}
0&1\\
1&0
\end{pmatrix}.\label{Sehsymmetry}
\end{equation}
At small excitation energies $|\varepsilon|\ll |M_{z}|,|\Delta|$ the $\varepsilon$-dependence of $S_{\rm in}$ may be neglected. Then Eq.\ \eqref{Sehsymmetry} together with unitarity ($S_{\rm in}^{-1}=S_{\rm in}^{\dagger}$) fully determine the scattering matrix,
\begin{equation}
S_{\textrm{in}}=\frac{1}{\sqrt{2}}\begin{pmatrix}
1&1\\
i&-i
\end{pmatrix}
\begin{pmatrix}
 e^{i \alpha}& 0\\
0&e^{-i \alpha}
\end{pmatrix},\label{eq:sin}
\end{equation}
up to a phase difference $\alpha$ between electron and hole  (which will drop out of the conductance and need not be further specified).

The scattering matrix $S_\textrm{out}$  for the conversion from Majorana modes to electron and hole modes can be obtained from $S_{\textrm{in}}$ by time reversal,

\begin{equation}
 S_\textrm{out}(\bm{M})=S_{\rm in}^{T}(-\bm{M})=\frac{1}{\sqrt{2}}
\begin{pmatrix}
 e^{i \alpha'}& 0\\
0&e^{-i \alpha'}
\end{pmatrix}
\begin{pmatrix}
1&i\\
1&-i
\end{pmatrix}.\label{eq:sout}
\end{equation}
The phase shift $\alpha'$ may be different from $\alpha$, because of the sign change of $\bm{M}$ upon time reversal, but it will also drop out of the conductance.

\begin{figure}[htb]
\centering
\includegraphics[width=0.9\linewidth]{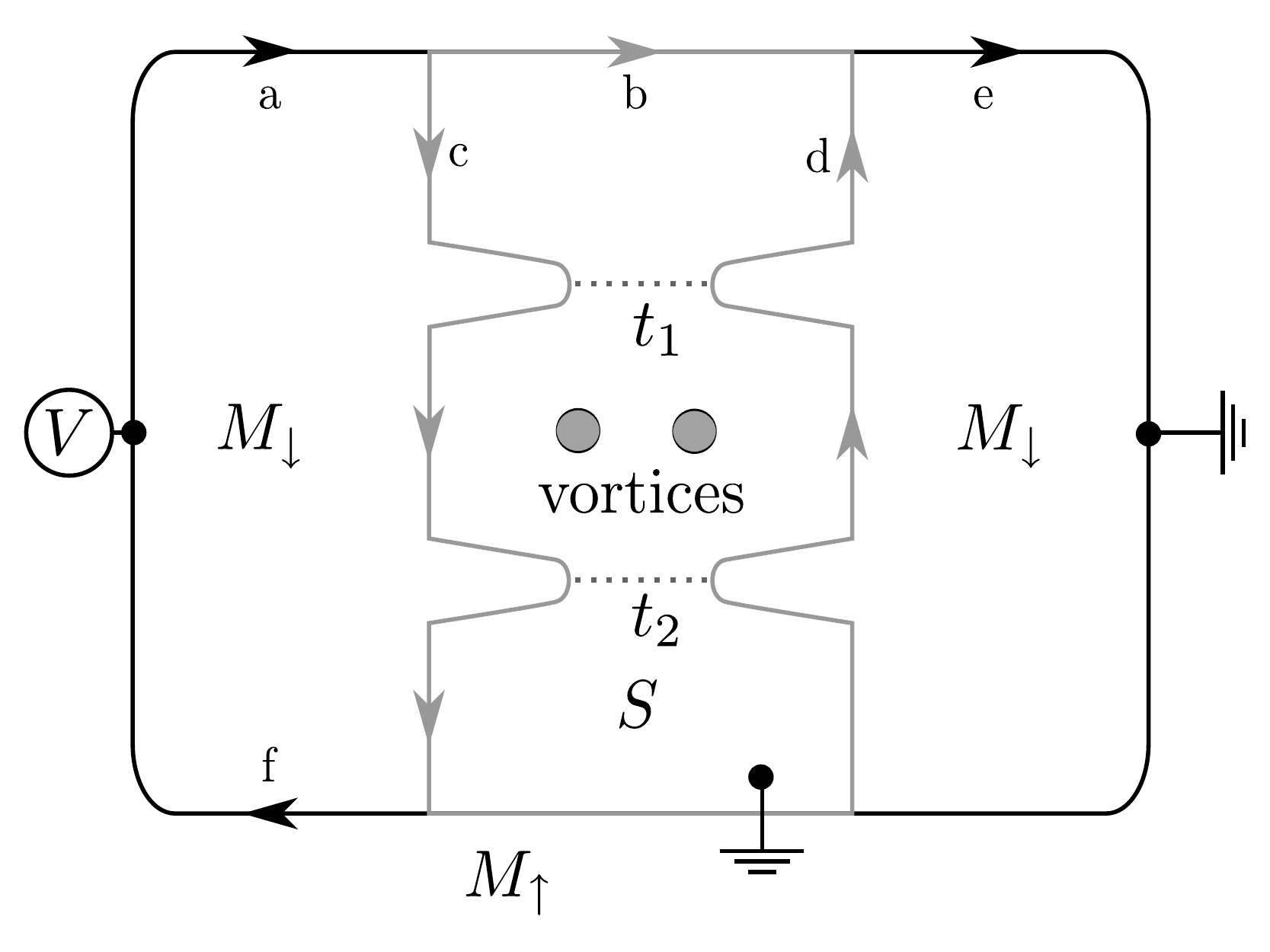}
\caption
{Fabry-Perot interferometer, allowing to measure the state of a qubit encoded in a pair of vortices. Black lines represent electron or hole modes at domain walls, gray lines represent Majorana modes at magnet-superconductor interface.
}
\label{fig:interference_setup}
\end{figure}

 The full scattering matrix $S$ of the Mach-Zehnder interferometer in Fig.~\ref{fig:setup_3d} is given by the matrix product
\begin{equation}
S\equiv\begin{pmatrix}
S_{ee}&S_{eh}\\
S_{he}&S_{hh}
\end{pmatrix}=
S_\textrm{out}\begin{pmatrix}
e^{i\beta_{b}}&0\\
0&e^{i\beta_{c}}
\end{pmatrix}
S_\textrm{in},\label{eq:totaltransfer}
\end{equation}
where $\beta_{b}$ and $\beta_{c}$ are the phase shifts accumulated by the Majorana modes along edge $b$ and $c$, respectively. The relative phase
\begin{equation}
\beta_{b}-\beta_{c}=\varepsilon\delta L/\hbar v_{m}+\pi+n_{v}\pi\label{betabbetac}
\end{equation}
consists of three terms: A dynamical phase (proportional to the length difference $\delta L=L_{b}-L_{c}$ of the two arms of the interferometer), a Berry phase of $\pi$ from the rotation of the spin-$1/2$, and an additional phase shift of $\pi$ per enclosed vortex.

The differential conductance follows from
\begin{equation}
G(V)=\frac{2e^2}{h}|S_{he}(eV)|^{2}=\frac{2 e^2}{h}\sin^2\left(\frac{n_{v}\pi}{2}+\frac{eV\delta L}{2\hbar v_m}\right).
\end{equation}
 As announced in the introduction, the linear response conductance $G(0)$ vanishes if the number of vortices is even, while it has the maximal value of $2e^{2}/h$ if the number is odd.

The Mach-Zehnder interferometer can distinguish between an even and an odd number $n_{v}$ of enclosed \textit{vortices}. The next step towards measurement based topological quantum computation is to distinguish between an even and an odd number $n_{f}$ of enclosed \textit{fermions}. If $n_{v}$ is odd, the parity of $n_{f}$ is undefined, but if $n_{v}$ is even, the parity of $n_{f}$ is a topologically protected quantity that determines the state of a qubit \cite{Nay08}. To electrically read out the state of a qubit encoded in a pair of charge-neutral vortices, we combine the Fabry-Perot interferometer of the FQHE \cite{Ste06,Bon06} with our Dirac-Majorana converter. 

The geometry is shown in Fig.~\ref{fig:interference_setup}. Electrons are injected in the upper left arm $a$ of the interferometer (biased at a voltage $V$) and the current $I$ is measured in the upper right arm $e$ (which is grounded). The electron at $a$ is split into a pair of Majorana fermions $\psi_{b}$ and $\psi_{c}$, according to the scattering matrix $S_{\rm in}$. A pair of constrictions allows tunneling from $\psi_{c}$ to $\psi_{d}$, with amplitude $t_{dc}$. Finally, the Majorana fermions $\psi_{d}$ and $\psi_{b}$ are recombined into an electron or hole at $e$, according to the scattering matrix $S_{\rm out}$. The resulting net current $I=(e^{2}/h)V(|T_{ee}|^{2}-|T_{he}|^{2})$ (electron current minus hole current) is obtained from the transfer matrix
\begin{equation}
T=S_{\rm out}\begin{pmatrix}
e^{i\beta_{b}}&0\\
0&t_{dc}\end{pmatrix}S_{\rm in}\Rightarrow I=\frac{e^{2}}{h}V\,\textrm{Re}\,\left(e^{-i\beta_{b}}t_{dc}\right).\label{TS}
\end{equation}
Notice that the current is proportional to the tunnel \textit{amplitude}, rather than to the tunnel probability. In the low-voltage limit, to which we will restrict ourselves in what follows, the phase shift $\beta_{b}$ vanishes and $t_{dc}$ is real (because of electron-hole symmetry) --- so $I$ directly measures the tunnel amplitude.

In general, two types of tunnel processes across a constriction contribute to $t_{dc}$: A Majorana fermion at the edge of the superconductor can tunnel through the superconducting gap to the opposite edge of the constriction either directly as a fermion or indirectly via vortex tunneling \cite{Fen07}. Fermion tunneling typically dominates over vortex tunneling, although quantum phase slips (and the associated vortex tunneling) might become appreciable in constrictions with a small capacitance \cite{Moo06}. Only vortex tunneling is sensitive to the fermion parity $n_{f}$, through the phase factor $(-1)^{n_{f}}$ acquired by a vortex that encircles $n_{f}$ fermions. Because of this sensitivity, vortex tunneling is potentially distinguishable on the background of more frequent fermion tunneling events.

The contribution to $t_{dc}$ from fermion tunneling is simply $t_{f,1}+(-1)^{n_v}t_{f,2}$, to lowest order in the fermion tunnel amplitudes $t_{f,1}$ and $t_{f,2}$ at the first and second constriction. There is no dependence on $n_{f}$, so we need not consider it further. 

To calculate the contribution to $t_{dc}$ from vortex tunneling, we apply the vortex tunnel Hamiltonian \cite{Fen07} $H_{i}=v_{i}\sigma_{i}\sigma'_{i}$, where $i=1,2$ labels the two constrictions and $v_{i}$ is the tunnel coupling. The operators $\sigma_{i}$ and $\sigma'_{i}$ create a vortex at the left and right end of constriction $i$, respectively. The lowest order contribution to $t_{dc}$ is of second order in the tunnel Hamiltonian, because two vortices need to tunnel in order to transfer a single Majorana fermion. The calculation of $t_{dc}$ will be presented elsewhere, but the $n_{v}$ and $n_{f}$ dependence can be obtained without any calculation, as follows.

Three terms can contribute to second order in $H_{i}$, depending on whether both vortices tunnel at constriction number 1 (amplitude $t_{1}^{2}$), both at constriction number 2 (amplitude $t_{2}^{2}$), or one at constriction number 1 and the other at constriction number 2 (amplitude $2t_{1}t_{2}$). The resulting expression for $t_{dc}$ is
\begin{equation}
t_{dc}=t_{1}^{2}+t_{2}^{2}+(-1)^{n_{f}}2t_{1}t_{2},\;\;\mbox{if $n_{v}$ is even}.\label{tdcresult1}
\end{equation}
We see that if the two constrictions are (nearly) identical, so $t_{1}\approx t_{2}\equiv t$, the tunnel amplitude $t_{dc}$ and hence the current $I_{\rm vortex}$ due to vortex tunneling vanish if the fermion parity is odd, while $I_{\rm vortex}=(e^{2}/h)V\times 4t^2$ if the fermion parity is even \cite{note1}.

In summary, we have proposed a method to convert a charged Dirac fermion into a pair of neutral Majorana fermions, encoding the charge degree of freedom in the relative phase of the two Majorana's. The conversion can be realized on the surface of a topological insulator at a junction between a magnetic domain wall (supporting a chiral charged mode) and two magnet-superconductor interfaces (each supporting a Majorana mode). We found that at low voltages the Dirac-Majorana conversion is geometry independent and fully determined by the electron-hole symmetry. It allows for the electrical read-out of a qubit encoded nonlocally in a pair of vortices, providing a building block for measurement based topological quantum computation.

We have benefited from discussions with C. L. Kane and B. J. Overbosch. This research was supported by the Dutch Science Foundation NWO/FOM.


\begin{thebibliography}{99}
\bibitem{Rad08} I. Radu, J. B. Miller, C. M. Marcus, M. A. Kastner, L. N. Pfeiffer, and K. W. West, Science \textbf{320}, 899 (2008). 
\bibitem{Dol08} M. Dolev, M. Heiblum, V. Umansky, A. Stern, and D. Mahalu, Nature \textbf{452}, 829 (2008). 
\bibitem{Wil09} R. L. Willett, L. N. Pfeiffer, and K. W. West, arXiv:0807.0221.
\bibitem{Moo91} G. Moore and N. Read, Nucl.\ Phys.\ B \textbf{360}, 362 (1991).
\bibitem{Nay08} C. Nayak, S. Simon, A. Stern, M. Freedman, and S. Das Sarma, Rev. Mod. Phys. \textbf{80} 1083 (2008).
\bibitem{Kit01} A. Yu. Kitaev, Phys. Usp. \textbf{44} (suppl.), 131 (2001).
\bibitem{Das05} S. Das Sarma, M. Freedman, and C. Nayak, Phys. Rev. Lett. \textbf{94}, 166802 (2005).   
\bibitem{Ste06} A. Stern and B. Halperin, Phys. Rev. Lett. \textbf{96}, 016802 (2006). 
\bibitem{Bon06} P. Bonderson, A. Kitaev, and K. Shtengel, Phys. Rev. Lett. \textbf{96}, 016803 (2006).
\bibitem{Bon08} P. Bonderson, M. Freedman, and C. Nayak, Ann. Phys. (New York) \textbf{324}, 787 (2008).
\bibitem{Gre92} M. Greiter, X. G. Wen, and F. Wilczek, Nucl. Phys. B \textbf{374}, 567 (1992).
\bibitem{Rea00} N. Read and D. Green, Phys. Rev. B \textbf{61}, 10267 (2000).
\bibitem{Iva01} D. A. Ivanov, Phys. Rev. Lett. \textbf{86}, 268 (2001).
\bibitem{Sar06} S. Das Sarma, C. Nayak, and S. Tewari, Phys. Rev. B \textbf{73}, 220502(R) (2006).
\bibitem{Tew07} S. Tewari, S. Das Sarma, C. Nayak, C. Zhang, and P. Zoller, Phys. Rev. Lett. \textbf{98}, 010506 (2007).
\bibitem{Fu08} L. Fu and C. L. Kane, Phys. Rev. Lett. \textbf{100}, 0964407 (2008); arXiv:0804.4469.
\bibitem{Hsi08} D. Hsieh, D. Qian, L. Wray, Y. Xia, Y. S. Hor, R. J. Cava, and M. Z. Hasan, Nature \textbf{452}, 970 (2008).
\bibitem{Zha08} H. Zhang, C.-X. Liu, X.-L. Qi, X. Dai, Z. Fang, and S.-C. Zhang, arXiv:0812.1622.
\bibitem{Xia08} Y. Xia, L. Wray, D. Qian, D. Hsieh, A. Pal, H. Lin, A. Bansil, D. Grauer, Y. S. Hor, R. J. Cava, and M. Z. Hasan, arXiv:0812.2078.
\bibitem{Sem07} G. W. Semenoff and P. Sodano, J. Phys. B \textbf{40}, 1479 (2007).
\bibitem{Bol07} C. J. Bolech and E. Demler, Phys. Rev. Lett. \textbf{98}, 237002 (2007).
\bibitem{Tew08} S. Tewari, C. Zhang, S. Das Sarma, C. Nayak, and D.-H. Lee, Phys. Rev. Lett. \textbf{100}, 027001 (2008).
\bibitem{Nil08} J. Nilsson, A. R. Akhmerov, and C. W. J. Beenakker, Phys. Rev. Lett. \textbf{101}, 120403 (2008).
\bibitem{Fen07}  P. Fendley, M. P. A. Fisher, and C. Nayak, Phys. Rev. B \textbf{75}, 045317 (2007); arXiv:0902.0998.
\bibitem{Moo06} J. E. Mooij and Yu. V. Nazarov, Nat. Phys. \textbf{2}, 169 (2006).
\bibitem{note1} Eq.\ \eqref{tdcresult1} assumes that the number $n_{v}$ of bulk vortices in between the two constrictions is even, so that $n_{f}$ is well-defined. When $n_{v}$ is odd, a vortex tunneling at constriction number 2 exchanges a fermion with the bulk vortices \cite{Iva01}. If both vortices tunnel at constriction number 2, the two fermion exchanges compensate with a phase factor of $-1$, but if one vortex tunnels at constriction 1 and the other at constriction 2, then the single fermion exchange prevents the transfer of a Majorana fermion across the superconductor. The resulting expression for $t_{dc}$ therefore contains only two terms, $t_{dc}=t_{1}^{2}-t_{2}^{2}$, if $n_{v}$ is odd.
\end{thebibliography}
\end{document}